\documentclass[aps,prb,twocolumn,showpacs,amsmath,amssymb,preprintnumbers,article]{revtex4}

\usepackage{color}
\usepackage{graphicx}

\begin {document}

\title{ Pressure driven topological phase transition in chalcopyrite ZnGeSb$_2$}

\author{Surasree Sadhukhan$^{1}$\footnote{both authors have equal contribution}, Banasree Sadhukhan$^{2, 3\ast}$, and Sudipta Kanungo$^1$}
\affiliation{$^1$ School of Physical Sciences, Indian Institute of Technology Goa, 403401 Ponda, India\\
$^2$ KTH Royal Institute of Technology, AlbaNova University Center, SE-10691 Stockholm, Sweden\\
$^3$ Institute for Theoretical Solid State Physics, IFW Dresden, Helmholtzstr. 20, 01069 Dresden, Germany}

\email{sudipta@iitgoa.ac.in,  b.sadhukhan@ifw-dresden.de}

\date{\today}

\begin{abstract}
Recently topologically non-trivial phases have been identified in few time-reversal invariant 
systems that lack of inversion symmetry. Using density functional theory based first-principles 
calculations, we report a strong topologically non-trivial phase in chalchopyrite ZnGeSb$_2$, which 
can act as a model system of strained HgTe. The calculations reveal the non-zero topological 
invariant ($Z_2$), the presence of Dirac cone crossing in the surface spectral functions with 
spin-momentum locking. We also show that the application of moderate hydrostatic 
pressure ($\sim$7 GPa) induces topological phase transition from topological non-trivial phase to a topologically trivial phase. A discontinuity in the tetragonal distortion of non-centrosymmetric ZnGeSb$_2$ plays a crucial role in driving this topological phase transition. 

\end{abstract}

\maketitle

\section{Introduction} 

Topological insulators (TIs) are the three-dimensional (3D) counterpart of two-dimensional (2D)-
quantum spin Hall insulators\citep{bernevig2006quantum,konig2007quantum,hsieh2008topological, 
zhang2009topological,xia2009observation}. They are characterized by a non-trivial topological 
Z$_2$ invariant associated to the bulk electronic structure 
\citep{qi2010quantum,hasan2010colloquium,moore2010birth,moore2007topological,fu2007topological, 
roy2009topological}. TIs have an insulating gap in the bulk band structure, but topologically 
protected gapless surface or edge states on the boundary. The surface states have an odd (even) 
number of mass-less Dirac cones associated with the Z$_2$ topological invariant which leads to 
strong (weak) TIs. They have inverted band ordering due to the switching of bands with 
opposite parity at high symmetry momenta of the Brillouin zone (BZ) around the Fermi level 
compared to their topological trivial phase.

Topological insulating phases have been theoretically predicted and experimentally observed in a 
variety of systems including HgTe/CdTe quantum well\citep{bernevig2006quantum,fu2007topological, 
konig2007quantum}, tetradymite semiconductors (Bi$_2$Se$_3$, Bi$_2$Te$_3$, Sb$_2$Te$_3$, Bi$_{1-x}$Se$_x$)\citep{zhang2009topological,xia2009observation,xia2009observation}, half-heusler 
compounds containing rare-earth elements Ln (Ln$=f^n$ lanthanides) like LnAuPb, LnPdBi, LnPtSb and 
LnPtBi\citep{chadov2010tunable,lin2010half,xiao2010half}. The half Heuslers TIs and Rare-earth carbides can host many 
interesting topological quantum phenomena properties by tuning number of strongly correlated 
$\textit{f}$ electrons\citep{canfield1991,goll2008,fisk1991,dzero2010,qi2008topological,
fu2008superconducting,fu2009probing,tanaka2009manipulation,yokoyama2010theoretical, ray2020tunable}.

In this context, it is of particular interest to identify materials for which the band ordering can 
be tuned, and topological properties can be switched via topological phase transitions. In principle, the phase transitions between topological trivial and non-trivial states can be achieved by the external 
perturbations such as the tuning the electro-negativity or tuning the strength of spin-orbit coupling (SOC) via substituting proper elements\citep{bernevig2006quantum,fu2007topological,konig2007quantum}, by alloying composition or chemical doping \citep{novak2015, singh2012, orlita2014} or by uniform hydrostatic pressure, and 
strain. Pressure induced topological phase transition have been identified in layered materials \cite{zhu2012}, 
BiTeI\citep{xi2013,facio2018strongly}, Pb$_{1-x}$Sn$_{x}$\cite{xi2014}, polar semiconductors BiTeBr \cite{ohmura2017}, topological crystalline insulators\cite{zhao2015}, rocksalt chalcogenides 
\cite{barone2013}, chalcopyrite compounds like CdGeSb$_2$ and CdSnSb$_2$\cite{juneja2018}.

Also ternary chalcopyrites of composition I-III-VI$_2$ and II-IV-V$_2$ provide an interesting platform to host topologically exotic phases\cite{feng2011}. Structurally these chalcopyrites can be viewed as strained HgTe which 
crystallizes in the zinc-blende lattice structure. Few of them can be realized as the topological 
insulator or Weyl semi metals in their native states\citep{feng2011,ruan2016}. Room temperature 
ferromagnetism can also be found in chalcopyrites by magnetic doping in addition to the topological 
non-trivial phases\citep{medvedkin2000, scho2002, erwin2004}. Very recently, photovoltaic phenomena related to the Berry phase of the constituting electronic bands were established in chalcopyrite 
compound ZnSnP$_2$\cite{sadhukhan2019first}. The wide availability of chalcopyrite semiconductors and the control of topological order via tuning lattice parameters, hydrostatic pressure, chemical doping opens new possibilities for 
this family ideal platform to further investigate the novel topological quantum phenomena.

The atomic positions (Wyckoff positions) in a materials depend on the symmetry of the systems followed by the space group of the  materials family. Chalcopyrite series compounds are such a materials where the Wyckoff positions are depend on the constituent atom in that materials. In this series of materials, each anion has two dissimilar cations as nearest neighbors. This leads to the structural distortion compared to ideal HgTe structure which is determined by both the lattice parameter's ratio and the internal displacement of the anions (anion-shift) towards one of the cation as compared to the ideal zinc-blende sites. So both the lattice parameters and atomic positions are important parameters to determine its ground state structure and consequently the properties whether it is in topological or trivial state.

Motivated by the above issues,  we focused here on a chalcopyrite compound ZnGeSb$_2$ which has a very low, positive band-inversion strength ($E_{\Gamma_6}-E_{\Gamma_8}$). This enables us to tune the topological band ordering via applying modest kind of perturbation in terms of uniform hydrostatic pressure of the order of $\approx$ 7 GPa.  Apparently, there are self contradicting reports found on ZnGeSb$_2$. It has been reported that the ZnGeSb$_2$ as a trivial insulator in their native state in the ref\cite{feng2011}. However there are contradictory report\cite{sreeparvathy2016} which shows some kind of massive Dirac states at the parent phase. But in the both studies the detailed study of the topological properties and quantities are still far from being investigation.

To clarify this discrepancy,  we have decided to address this apparent contradiction through our detailed electronic structure investigation of the topological quantities. We cross checked all our results in two different basis sets to established our results on ZnGeSb$_2$ firmly. We did a similar investigation with another Chalcoyrite compound ZnSnP$_2$ in our previous report\cite{sadhukhan2019first}. Our fully relativistic\cite{eschrig2004} calculations using in a four component Dirac formulation, however reveal that the ZnGeSb$_2$ is in a topologically non-trivial state at the parent phase with non-zero Z$_2$ indices and non-trivial surface states crossing near the Fermi energy with the spin-momentum locking and undergoes a topological phase transition to trivial insulating phase at high pressure ($\approx$ 7 GPa). Here in our studies we explicitly correlate the topological phase transition with the position of the anion (Sb atom) and 
on the structural distortions in ZnGeSb$_2$.

\section{Calculations Methodology} 
\label{method}
Density functional theory (DFT) based electronic structure calculations were performed using the local orbital basis set in the full potential framework as implemented in the FPLO code\cite{fplo1,fplo2} as well as for structural aspects through the plane-wave basis set based on a pseudo-potential framework as incorporated in the Vienna $\textit{ab-initio}$ simulation package (VASP)\citep{vasp1,vasp2}. The generalized gradient exchange-correlation approximated (GGA) functional was employed following the Perdew–Burke–Ernzerhof (PBE) prescription\cite{PBE}. For the plane-wave basis, a 600 eV cut-off was applied. The structural optimization were performed by relaxing the both \textit{a} and \textit{c} lattice parameters and the internal atomic positions of all the atoms toward the equilibrium until the Hellmann–Feynman force becomes less than 0.001 eV/$\AA$. We also cross checked the obtained optimized lattice parameters through the energy minimization method as shown in the appendix Fig.\ref{fig:app1}. A k-point mesh of 6 $\times$ 6 $\times$ 4 in the Brillouin zone was used for the calculations. All the structures are optimized and  the self-consistent calculations are converged  with tight convergence threshold for the energy ($10^{-7} eV$).

The tight-binding model is constructed with the Wannier function basis set to investigate its topological properties. Wannierization is an energy selective method that produces the low energy, few orbital Hamiltonian defined in the effective Wannier function basis by integrating out the degrees of freedom that are not of interest through re-normalization method. In the downfolding calculations, we have kept only Sb-$s$, $p$ and Ge-$p$ orbitals as an active degrees of freedom in the projection and downfolded all other orbitals. The Wannier interpolated bands, $Z_2$ invariant, Wannier charge centers (WCC), surface states and the spin textures were calculated using Wannier90\citep{wannier1,wannier2, wannier3} and WannierTool\cite{wanniertool} starting from full DFT calculations. The reliability of the calculations and the results has been cross checked in the other basis set using full potential method as implemented in FPLO\cite{fplo1,fplo2}. We also calculate the Z$_2$ invariant using Wannier-Center algorithm within FPLO (Full-potential local-orbital minimum-basis)\cite{fplo1,fplo2}. The obtained electronic structure results from two different basis sets are consistent with each others, confirms the robustness of our analysis. 

\section{Crystal Structure and symmetry}
\begin{figure*}
\includegraphics[width=18cm]{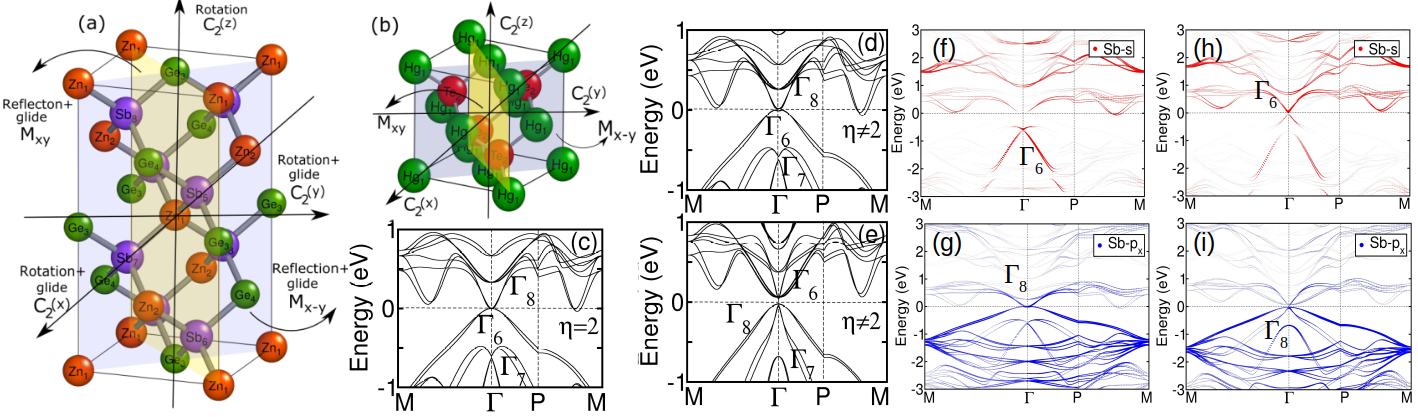}
\caption{(color online) Unit cell of the (a) body centered tetragonal ZnGeSb$_2$ and (b) HgTe. The associated symmetry axes and planes are also shown by line and shade respectively. Calculated GGA+SOC band structures along the high symmetry line in the BZ with (c) artificial structure of ZnGeSb$_2$, where $\eta = 2$, with ${\delta}u = 0$, (d) and (e) with real distorted tetragonal crystal of ZnGeSb$_2$ for ambient structure and high pressure structure respectively. (f)-(i) show the calculated GGA+SOC orbital projected band structure of Sb atoms. (f) and (g) represent parent phase (topological non-trivial), (h) and (i) represent high pressure (topological trivial) phases of ZnGeSb$_2$ respectively for the the projected bands of Sb-$5s$, Sb-$5p_x$. The bands for the Sb-$5p_y$, and Sb-$5p_z$ orbitals are similar of Sb-$5p_x$ for in both phases, shown in the appendix \ref{fig:app2}. $\Gamma_6$ and $\Gamma_8$ band characters are marked in the figure and the band inversion in topological phase happens between $5s$ ($\Gamma_6$) and $5p$ ($\Gamma_8$) orbitals at the $\Gamma$ point in the BZ.  Fermi energy set at the zero in energy axis.}
\label{fig:Fig1}
\end{figure*}

ZnGeSb$_2$ crystallize in the body-centered tetragonal structure which in the chalcopyrite phase has a space group I$\bar{4}$2d (no. 122). Structurally it is a superlattice of zinc-blende structure, like ideal HgTe, doubling the unit cell along the crystallographic \textit{z} direction. In the ZnGeSb$_2$, each anion Sb has
two Zn and two Ge cations as nearest neighbors as shown in Fig.\ref{fig:Fig1}(a). Due to dissimilar 
atoms as neighbours the anion acquires an equilibrium position closer to one pair of cation than to 
other. The wyckoff positions of the different atoms in the tetragonal unit cell are: Zn atom at (0, 0, 0); Ge atom at (0, 0, 0.5) and Sb atom at ($x$, 0.25, 0.125), where $x$ is the anion displacement parameter. The full structural optimization calculations and the energy minimization of the structural parameters (shown in Fig. \ref{fig:app1}) gives us the lattice parameters \textit{a}=6.20 $\AA$, \textit{c}=12.33 $\AA$ and the anion displacement parameter turn out to be \textit{x}=0.2454. Since, this compound is not synthesized experimentally yet, its very important to examine the thermodynamic and the dynamical stability. The calculated phonon density of states with absence of any negative frequency phonon mode and the Helmholtz free energy, as shown in the Fig. \ref{phonon} confirm that the stability of the structures. The internal displacement of the anion in chalcopyrite compounds of general form ABC$_2$ is defined as ${\delta}u = {(R^2_{AC}-R^2_{BC})}/{a^2}$ where R$_{AC}$ and R$_{BC}$ are the bond lengths between the anion C and its two nearest A and B cations\cite{ruan2016} respectively. In the most general case of chalcopyrite, x$\neq$ 0.25 and $\eta$ $\neq$ 2, where $\eta$=c/a, a and c are the lattice parameters. ${\delta}u$ the measures the tetragonal distortion and responsible for the breaking of the inversion symmetry in ZnGeSb$_2$. The non-centrosymmetric ZnGeSb$_2$ crystal has two twofold glide rotational symmetries along C$_2$(x) and C$_2$(y) respectively and two glide mirror symmetry axes M$_{xy}$ and M$_{x-y}$, in addition to that of the pure twofold rotational symmetry along C$_2$(z).

The binary compound with zinc-blende structure (cubic symmetry), $x$ is 0.25 and $\eta = c/a=1$. The zinc-blende HgTe superlattice can be regarded as a model chalcopyrite like ABC$_2$ with A=B=Hg, C=Te, $\eta=2$, and ${\delta}u =0$. Interestingly, in the chalcopyrite, the cubic symmetry is broken due to the tetragonal distortion ($\eta \neq 1$) and the internal displacement (${\delta}u \neq 0$). Therefore, the ZnGeSb$_2$ can be though of as the strained HgTe (see Fig. \ref{fig:Fig1}(b)).

\section{Electronic structure and band topology} 

The calculated electronic band structures with SOC are shown in Fig.\ref{fig:Fig1}(c)-(e) along the high symmetry lines in the BZ. In an ideal zinc-blende structure with cubic symmetry as found in HgTe (${\delta}u = 0$, $x=0.25$, $\eta = 1$), the $p$ orbital symmetric four fold degenerate $\Gamma_8$ states of total angular momentum $J = \frac{3}{2}$, lies above the $s$ orbital symmetric two fold degenerate $\Gamma_6$ and $\Gamma_7$  states of total angular momentum of $J = \frac{1}{2}$. To understand the band topology relating to breaking of symmetry and tetragonal distortion, we did a model calculations, by considering a hypothetical structure, where $x$=0.25 for the Sb coordinate and setting $\eta$=2 ($\eta =c/a$, where, $c$ and $a$ are the exact lattice parameters of ZnGeSb$_2$ obtained through the structural 
optimization in the ambient structure) as shown in Fig.\ref{fig:Fig1}(c). In this hypothetical structure, ${\delta}u=0$, i.e without tetragonal distortion and the same space group keeps the constant point group symmetry of the structure. The fourfold degeneracy of the $\Gamma_8$ is protected when $\eta$=2. The Dirac cone type
feature at $\Gamma$ point at the Fermi energy, similar to that of the case of HgTe with $\eta = 2$. However, the ZnGeSb$_2$ formed in a chalcopyrite structure of tetragonal unit cell where $\eta \neq$ 2 and x $\neq$ 0.25, the four fold degenerate light-hole and heavy-hole subbands of $\Gamma_8$ states was lifted and form typically the top set of the valence bands and the bottom set of the conduction bands at $\Gamma$ point above Fermi energy as shown in Fig.\ref{fig:Fig1}(d). In ZnGeSb$_2$, $E_{\Gamma_6}-E_{\Gamma_8} < 0 $, which indicates its topological nature in the native ambient structure phase.

\begin{figure}
\begin{center}
\includegraphics[width=9cm]{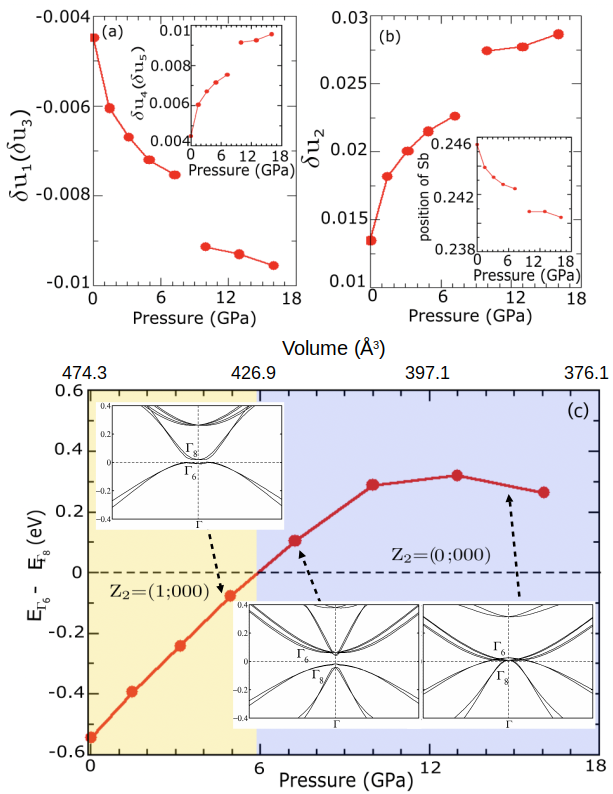}
\end{center}
\caption{(color online) (a) and (b) show the variation of the tetragonal distortions ${\delta}u's$ as a function of applied external hydrostatic pressure. The inset in (b) represents the deviation of Sb atomic position from ideal $x=0.25$ as a function of external pressure. Structural data are calculated via structural optimization at the DFT level. $\delta{u_1} = \frac{(Zn_1-Sb_5)^2-(Ge_3-Sb_5)^2}{a^2}$,  $\delta{u_2} = \frac{(Zn_2-Sb_5)^2-(Ge_4-Sb_6)^2}{a^2}$ and $\delta{u_3} = \frac{(Zn_2-Sb_8)^2-(Ge_4-Sb_7)^2}{a^2}$.  $\delta{u_4} = \frac{(Zn_2-Sb_5)^2-(Ge_4-Sb_7)^2}{a^2}$ and $\delta{u_5} = \frac{(Zn_2-Sb_8)^2-(Ge_4-Sb_6)^2}{a^2}$. (c) The calculated $E_{\Gamma_6}-E_{\Gamma_8}$ and Z$_2$ invariant quantities ($\gamma$; $\gamma_1\gamma_2\gamma_3$) of ZnGeSb$_2$ as a function of applied hydrostatic pressure (and corresponding volume of the unit cell) show the topological phase transition at $\approx$ 6-7 GPa. It is in topological non-trivial phase for E$_{\Gamma_6}$-E$_{\Gamma_8} < 0 $ and topological trivial phase for $E_{\Gamma_6}-E_{\Gamma_8} > 0 $. Insets show the $\Gamma_6$ and $\Gamma_8$ bands positioning before and after phase transition.}
\label{fig:Fig2}
\end{figure}

Since, the local gap at the $\Gamma$ point between the $\Gamma_6$ and $\Gamma_8$ band is very tiny, any moderate level of external perturbation may flip the band ordering at the $\Gamma$ point in the BZ. With this motivation, we applied uniform hydrostatic pressure to the parent structure of ZnGeSb$_2$ keeping the space group symmetry (I$\bar{4}$2d) unaltered. The calculated band structure in presence of SOC, showed in the Fig.\ref{fig:Fig1}(e). We found that at about moderate uniform hydrostatic pressure ($\approx$ 7 GPa), the topological band ordering at the $\Gamma$ point in the BZ gets flipped, i.e after applying pressure, $\Gamma_6$ and $\Gamma_8$ forms the bottom of the conduction band and top of the valance band respectively, making the $E_{\Gamma_6}-E_{\Gamma_8} > 0$ as shown in the Fig.\ref{fig:Fig2}(c). Therefore, ZnGeSb$_2$ undergoes topological phase transition from topological non-trivial to topological trivial phase by flipping of $\Gamma_6$ and $\Gamma_8$ band ordering as a function of applied external uniform hydrostatic pressure like perturbation.

To understand the evaluation of band structure across the topological phase transition mediated 
through hydrostatic pressure, we compared the orbital projected band structures for the parent 
(topologically non-trivial) and high pressure (topologically trivial) phase of ZnGeSb$_2$ in the 
Fig.\ref{fig:Fig1} (f)-(i). The calculations show that the only Sb states are dominating near the 
Fermi energy, with small contribution of Ge states, therefore  we have shown the orbital 
contribution of Sb atoms of ZnGeSb$_2$ for both parent and high pressure phases. In the parent 
structure phase, at the $\Gamma$ point of the BZ, the valence band maximum (VBM) has the major 
contribution from Sb-$5s$ orbital, forming the $\Gamma_6$ bands, whereas the conduction band minimum (CBM) has the major contribution from Sb-$5p_x$, Sb-$5p_y$ and a minor contribution from  Sb-$5p_z$ orbitals forming the $\Gamma_8$ bands. However, on the contrary in the high pressure phase, the VBM and CBM are formed by the Sb-$5p$ and Sb-$5s$ orbitals respectively, which suggests that, at the $\Gamma$ point the bands get inverted compared to the parent phase. Therefore, from our calculations, it confirms that the relative weightage of the Sb-$5s$ and Sb-$5p$ are flipped between the parent ambient phase and high pressure structure phase, leading to a topological phase transition between the parent ambient topologically non-trivial phase (where $E_{\Gamma_6}-E_{\Gamma_8} < 0 $ ) to the high pressure topologically trivial phase (where $E_{\Gamma_6}-E_{\Gamma_8} > 0 $ ), which is exactly similar to that of the HgTe/CdTe quantum well \citep{bernevig2006quantum,fu2007topological, konig2007quantum}.

The application of the hydrostatic pressure leads to the topological phase transition, which can be 
understood from the evolution of the tetragonal distortion. Under the application of the hydrostatic
pressure, the structural tetragonal distortion decreases which is quantify by ${\delta}u$ 
terms. The bond lengths and corresponding ${\delta}u$'s are shown in the Fig.\ref{fig:Fig2}(a)-(b).
The ${\delta}u$ decreases i.e the strength of tetragonal distortion decreases if we increase pressure. The distortions $\delta{u_1}$ shows monotonic fall and $\delta{u_2}$ and $\delta{u_3}$ show monotonic raise with pressure respectively, however have a discontinuous jump in $\delta{u}$'s at the same pressure range where the band ordering of $\Gamma_8$-$\Gamma_6$ switches i.e at the topological phase transition point. We also found that this increase and jump in the distortions is nicely corresponds to the deviation of the Sb atom positions from the ideal value of $x=0.25$ in the ZnGeSb$_2$ in the body centered I$\bar{4}$2d structure, as clearly shown in the inset of Fig.\ref{fig:Fig2}(b).

\section{topological invariant and surface Dirac cones} 

To ensure the topological character and phase transition, we calculated $Z_2$ invariant quantities 
for the both parent ambient and high pressure phases. The corresponding energy difference between 
$\Gamma_6$ and $\Gamma_8$ bands ($E_{\Gamma_6}-E_{\Gamma_8}$) as a function of external uniform hydrostatic pressure as shown in Fig.\ref{fig:Fig2}(c). The light-hole and heavy-hole subbands of $\Gamma_8$ symmetry is separated by a small local energy gap at $\Gamma$ point. Therefore, $Z_2$ topological invariant can still be defined for the valence bands as they are separated from the 
conduction bands by local energy gap in the BZ zone around $\Gamma$ point. We found that the 
$E_{\Gamma_6}-E_{\Gamma_8}$ is $< 0$ (top inset Fig.\ref{fig:Fig2}(c)) and Z$_2$ invariant quantities ($\gamma$; $\gamma_1\gamma_2\gamma_3$) are (1;000) up to the pressure $\approx$ 6-7 GPa, which indicates its strong topological nature ($\gamma$ is the strong topological index and $\gamma_i$ are the three weak topological indices). However, as the pressure increases, the major $s$-like $\Gamma_6$ symmetry bands rise above the $\Gamma_8$ bands leads to the $E_{\Gamma_6}-E_{\Gamma_8}$ $> 0$ (see bottom inset Fig.\ref{fig:Fig2}(c)).  

The Z$_2$ invariant quantities merged to the topologically trivial value of (0;000), which indicates that ZnGeSb$_2$ became topologically trivial phase by application of moderate hydrostatic pressure. However, increasing pressure further above $\approx$ 10 GPa, the linear trends of $E_{\Gamma_6}-E_{\Gamma_8}$ follow a downturn. This is associated by the crossing of top of the valence band ($\Gamma_8$) above the Fermi level (see bottom inset of Fig.\ref{fig:Fig2}(c)).

\begin{figure*}[t!]
\includegraphics[width=17cm]{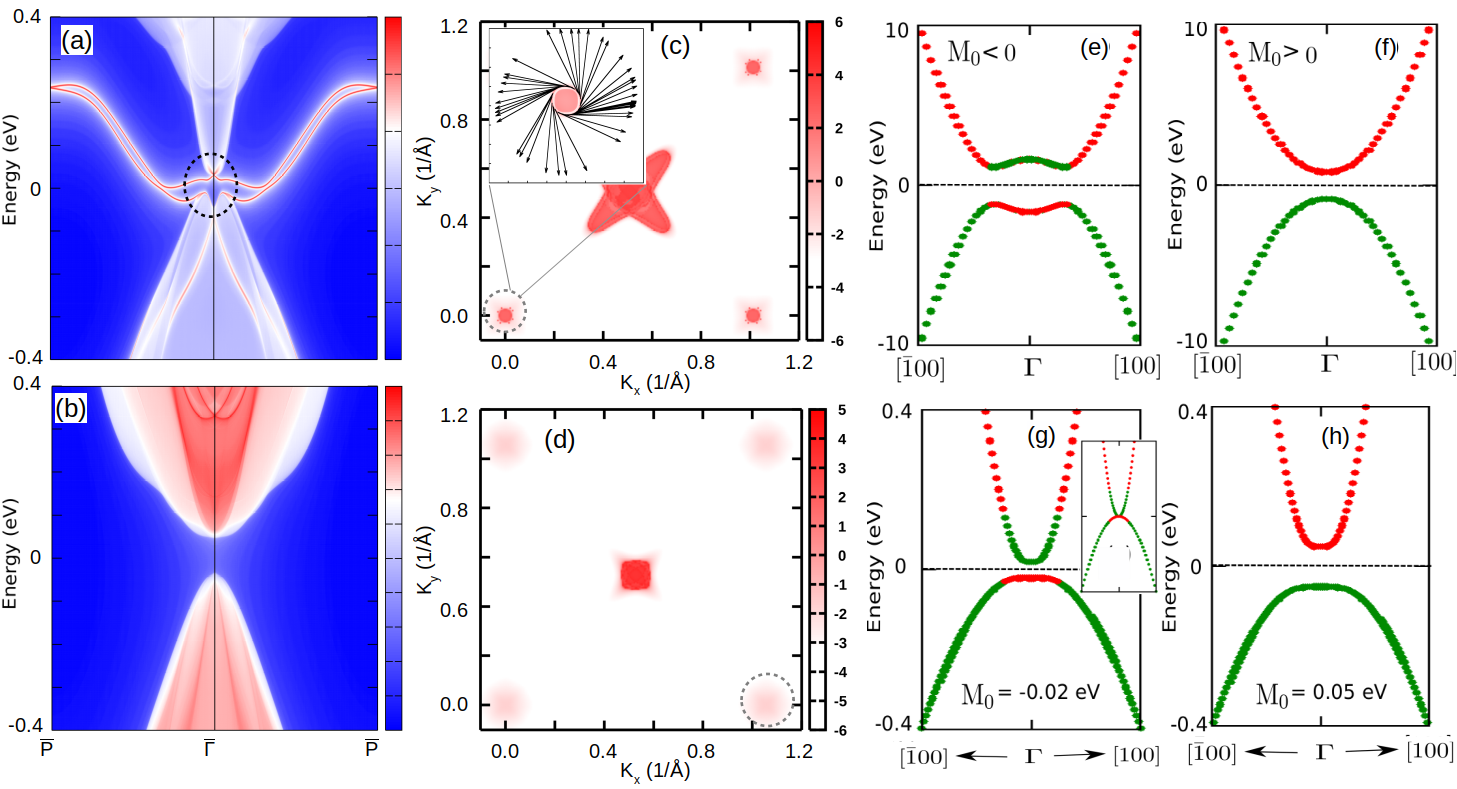}
\caption{(color online) The surface spectral distribution along the surface BZ
for topological (a) non-trivial (ambient pressure) (b) trivial phases (high pressure). Zero in the energy scale is set at the Fermi energy. (c) and (d) show the calculated Berry curvatures and the spin textures for topological non-trivial (parent phase) and trivial phases (high pressure) respectively. Band dispersion obtained from $H{\bf (k)}$ for (e) M$_0$ $<0$ which is in topological phase and (f) M$_0$ $> 0$ which is in non-topological phase. ZnGeSb$_2$ (g) $M_0 = -0.02$ eV and (h) $M_0 = 0.05$ eV. The inset of figure (g) corresponds to $M_0 = 0 $. }
\label{fig:Fig3}
\end{figure*}

Interesting point to be noted that, this topological phase transition in ZnGeSb$_2$ from topological
non-trivial to trivial phases are not due to the breaking of time reversal symmetry. The first hypothetical band structure (Fig.\ref{fig:Fig1} (c)) actually mimicking the two times supercell of 
an ideal cubic HgTe system. By reducing the symmetry of these hypothetical structure further, by 
setting $\eta \neq 2$ and $x \neq 0.25 $, we eventually have the actual structure of ZnGeSb$_2$ 
(figure \ref{fig:Fig1} (d)), which has substantial tetragonal distortions as defined in terms of 
${\delta}u$'s.  Point to be noted that, the calculated Z$_2$ invariant quantities 
(1;000)($\gamma$; $\gamma_1\gamma_2\gamma_3$) for both the hypothetical band structures and the original band structure of ZnGeSb$_2$ are topologically non-trivial and the reduction of symmetry does not mess up with the topological Z$_2$ invariant characters.

In parent state of ZnGeSb$_2$, the $K_x=0.5$, $K_y=0.5$ and $K_z=0.5$ planes have zero $Z_2$ invariant, while $K_z=0$ plane is non-trivial as shown in appendix Fig.\ref{fig:app4}. For $K_z=0$ plane, the WCC crosses the reference line for odd no of times, which results in $Z_2=1$. So the topological indices for ZnGeSb$_2$ are (1;000). Here $K_z=0$ $\neq$ $K_z=0.5$ $\rightarrow$ $\nu_0$ = 1, $\nu_1$=$Z_2$($K_x=0.5$)=0, $\nu_2$=$Z_2$($K_y=0.5$)=0, $\nu_3$=$Z_2$($K_z=0.5$)=0 as shown in appendix Fig.\ref{fig:app4}). Under hydrostatic pressure, all $K_z=0$, $K_x=0.5$, $K_y=0.5$ and $K_z=0.5$ planes have zero $Z_2$ invariant. So the topological indices for ZnGeSb$_2$ under pressure are (0;000), as shown in the appendix Fig.\ref{fig:app4}.

To elucidate our results, we also calculated the spectral distribution at the surface of the semi-infinite slab of ZnGeSb$_2$ via Green’s function techniques as implemented within the 
WannierTool\cite{wanniertool}. The tight binding model has been extracted using maximally projected Wannier functions (WFs). The surface states (SS) emerging from the conduction band (CB) and valence band (VB) exhibits Dirac like crossing at the $\bar{\Gamma}$ point in the BZ as shown in Fig.\ref{fig:Fig3}(a), which further ensures the strong topological character in ZnGeSb$_2$.  We also calculated the spectral distribution at the high pressure (7.22 GPa), as shown in Fig.
\ref{fig:Fig3}(b), where there are clear gap between valence and conduction bands near the Fermi 
energy and absence of any surface states, which further ensure topologically trivial band insulating
phase, at the high pressure. This establishes the topological and trivial character in the parent and high pressure phase in ZnGeSb$_2$. To verify the surface Dirac cone, we also calculated the spectral distribution at the surface time reversal invariant momenta (TRIM) points scanning the whole BZ in ZnGeSb$_2$ via Green’s function techniques as implemented within PYFPLO module \cite{fplo1,fplo2}. This establishes the strong topological character as surface states have an odd number of Dirac crossing in the parent phase which gone away at high pressure (see Fig.\ref{fig:app5} in the appendix). This ensure the topological phase transition at high pressure in ZnGeSb$_2$.

The calculated Berry phase for the parent phase of ZnGeSb$_2$ is very close to 2$\pi$ (1.995$\pi$), which drops down close to zero (0.010$\pi$) at about 7 GPa, which suggest that there are a topological phase transition from the topological non-trivial phase at parent phase to the topological trivial band insulating phase at the high pressure. To characterize the topological phase in more detail we also have calculated the Berry curvature and spin texture for the parent  and high pressure phases as shown in Fig.\ref{fig:Fig3}(c) and Fig.\ref{fig:Fig3}(d) respectively. The figures clearly show the changes in Berry curvature by applying the pressure. Moreover, we also found the existence of the spin-momentum locking signature (inset of Fig.\ref{fig:Fig3}(c)) in the spin texture at the parent phase and absence of it in the high pressure phase, also confirming the topological non-trivial phase to trivial phases transition by application of pressure. 

\section{Low-energy effective model} 
The topological nature and its effect under hydrostatic pressure in ZnGeSb$_2$ is determined by the physics near the $\Gamma$ point.  Therefore, we have constructed a low-energy effective $\bf{k.p}$ model Hamiltonian considering time reversal ($\tau$) and spatial (C$_2$(x), C$_2$(y), C$_2$(z) and 
M$_{xy}$) symmetries\cite{varjas2018qsymm} in ZnGeSb$2$. The $4 \times 4$ Hamiltonian is 
constructed in the basis of $\vert$ $\Gamma_6$, $\Gamma_8$;$\pm$ 1/2$\rangle$=$\vert$ s,$\uparrow\ 
\downarrow$ $\rangle$ and $\vert$ $\Gamma_8$;$\pm$ 3/2$\rangle$=$\vert$($p_x$ $\pm$ i$p_y$, 
$\uparrow\downarrow$ $\rangle$ of Sb atom. In such a basis set, the Hamiltonian can be written as 
\begin{eqnarray}
    &&H({\bf k})=\left(
    \begin{array}{cccc}
        \mathcal{M}({\bf k})&\mathcal{A}({\bf k})&0&-\mathcal{B^*}({\bf k})\\
        \mathcal{A}({\bf k})&-\mathcal{M}({\bf k})&\mathcal{B^*}({\bf k})&0\\
        0&\mathcal{B}({\bf k})&\mathcal{M}({\bf k})&\mathcal{A}({\bf k})\\
        -\mathcal{B}({\bf k})&0&\mathcal{A}({\bf k})&-\mathcal{M}({\bf k})
    \end{array}
    \right)
    \label{eq:Heff}
\end{eqnarray}
with $\mathcal{M}({\bf k})=M_0-M_1k_z^2-M_2k_\perp^2$, $\mathcal{A}({\bf k})=A(k_z^2-k_{+}k_{-})$, 
$\mathcal{B}({\bf k})={\beta}{k_+}k_z$, $k_\pm=k_x\pm ik_y$. So the parameters of the low energy 
effective model are M$_{012}$, A and $\beta$ in which M$_0$ determines the topology of band 
structure. This is equivalent to Dirac mass parameter\citep{konig2007quantum, 
sklyadneva2016pressure} and corresponds to the energy difference between basis orbitals. Other 
parameters are kept fixed in our calculation to reproduce the topological transitions (M$_1$=1.0 
eV${\AA}^2$, M$_2$=-1.0 eV${\AA}^2$, A=1.0 eV${\AA}^2$ and $\beta$=1.0 eV${\AA}^2$). The system is in topological insulating phase when M$_0$M$_1$ $<0$ and in normal insulating phase when M$_0$M$_1$ $>0$ as shown in Fig.\ref{fig:Fig3}(e)-(f). M$_0$M$_1$ $=0$ equivalent to mass less Dirac semi metals though the bands have a different physical interpretation.

These situations are similar to the topological phase transition under hydrostatic pressure 
presented above in Fig.\ref{fig:Fig3}(e)-(h). Such a transition is driven by the sign change of the mass term (M$_0$) in the three-dimensional Dirac equation which is the theory of the topological phase transition between the topological non-trivial and trivial insulators. The topological phase in ZnGeSb$_2$ occurs in the inverted regime where $E_{\Gamma_6}-E_{\Gamma_8} < 0$ and Z$_2$=(1;000), then M$_0 < 0$. The topological band ordering can be continuously tuned via moderate hydrostatic pressure until $E_{\Gamma_6}-E_{\Gamma_8} > 0$ and Z$_2$=(0;000). Then the mass term M$_0$ is getting flipped from a negative to positive value i.e, M$_0 > 0$ (see Fig.\ref{fig:Fig3}(g)-(h)).  This corresponds to a phase transition from topological metallic state to normal insulating phase in ZnGeSb$_2$.

\section{Discussion and Conclusion} 
ZnGeSb$_2$ is a strained HgTe whose topological band ordering can be tuned 
for appropriate applications via a moderate perturbation. The parent phase of ZnGeSb$_2$ is a 
strong topological non-trivial state (Z$_2$=(1;000)) with odd number surface states crossing in the 
surface BZ.  Apparently our findings are contradicting with the results showed in the previous 
report\cite{feng2011}. Points to be noted here are that, for the chalcopyrite class of materials,   both the in-plane lattice parameter \textit{a} and the out of plane lattice parameter \textit{c} are crucial along with the anion displacement parameter i.e the wyckoff coordinate (\textit{x}) of the Sb atom in determining the physical properties. We found that authors in the ref\cite{feng2011} showed the $E_{\Gamma_6}-E_{\Gamma_8}$ as a function of the lattice parameter \textit{a} only, without mentioning anything about the lattice parameter \textit{c} or the free wyckoff coordinate (\textit{x}) of the Sb atom in their calculations. The obtained in-plane lattice parameter  \textit{a} in ref\cite{feng2011}, is in quite good agreement with the value obtained by our calculations, however, we also mentioned the lattice parameter \textit{c} and free wyckoff coordinate (\textit{x}) of the Sb atom obtained through our structural optimization and the energy minimization as shown in appendix Fig.\ref{fig:app1}.  We also found that, keeping the lattice parameter \textit{a} same as it obtained in the energy minimization, if we only change the lattice parameter \textit{c} by around 10$\%$, then we found that the $E_{\Gamma_6}-E_{\Gamma_8} > 0$ i.e topological trivial state, as shown in the appendix Fig.\ref{fig:app3}, which in turn confirms that determination of out of plane lattice parameter \textit{c} is also very crucial in determining the topological properties. On the other hand, the topological nature of the ground state does not switch if we independently vary the in-plane lattice parameter \textit{a} or the free wyckoff coordinate (\textit{x}) of the Sb atom keeping other structural parameters same as that of the equilibrium value. Our obtained results have been cross checked in two different basis sets to establish our results on ZnGeSb$_2$ firmly.

By application of moderate uniform hydrostatic pressure ($\approx$ 7 GPa), 
ZnGeSb$_2$ undergoes a topological phase transition to the topologically trivial band insulating 
phase followed by normal metal. The topological phase transition is also associated with the 
tetragonal distortions (${\delta}u's$) in ZnGeSb$_2$. With increasing the pressure, the structural 
distortion parameters (${\delta}u's$) are decreasing and at above a critical value there is a jump 
in the ${\delta}u$. The topological phase transition is associated with the change in the wave 
function parity of $\Gamma_6-\Gamma_8$ band and sign of the mass term at the $\Gamma$ point in the 
BZ caused by the structural distortion. We hopped that our study will extend to any compounds in 
this chalcopyrite series. Therefore, it is interesting to investigate further the topological 
properties in others for realizing topological devices for quantum computing.

\section{Acknowledgments}
SS acknowledged IIT Goa (MHRD, Govt. of India) for providing fellowship. BS thanks Jeroen van den Brink for fruitful discussions and Ulrike Nitzsche for technical assistance with the computational resources in IFW cluster. Author SK thanks Department of Science and Technology (DST), Govt. of India for providing INSPIRE research funding (Grant No. DST/INSPIRE/04/2016/000431; IFA16-MS91).

\appendix

\section{Energy minimization calculations of structural parameters}
\begin{figure}[t!]
\includegraphics[width=6cm]{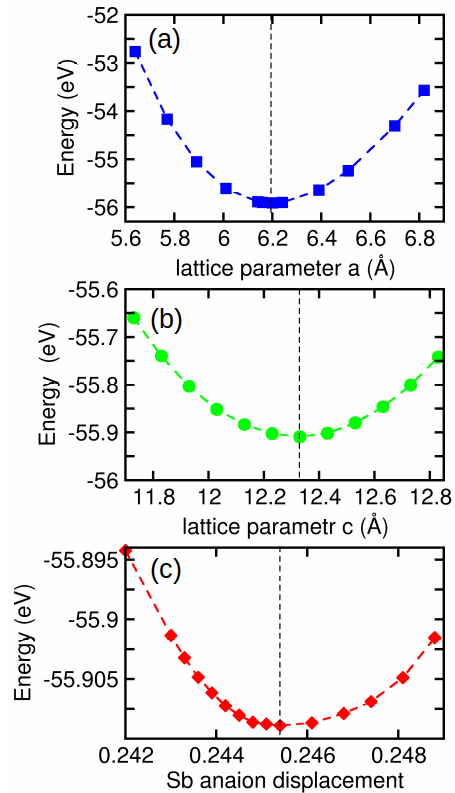}
\caption{(Color online) Total energy calculations as a function of the lattice parameters (a) \textit{a} and (b) \textit{c}  respectively. (c) Total energy calculations as a function of the wyckoff free coordinate \textit{x} of the Sb atoms. Energetically minimized values are marked by the dotted vertical lines in each plots.}
\label{fig:app1}
\end{figure}

Energy minimization of structural parameters viz. lattice parameters \textit{a}, \textit{c} and free wyckoff position of the Sb atoms as described in Sec.\ref{method}, shown in Fig. \ref{fig:app1}. The detail first principles calculations are done with this optimized and energy minimized structural parameters.

\section{Thermodynamic and dynamical stability}
\begin{figure}[t!]
\includegraphics[width=7cm]{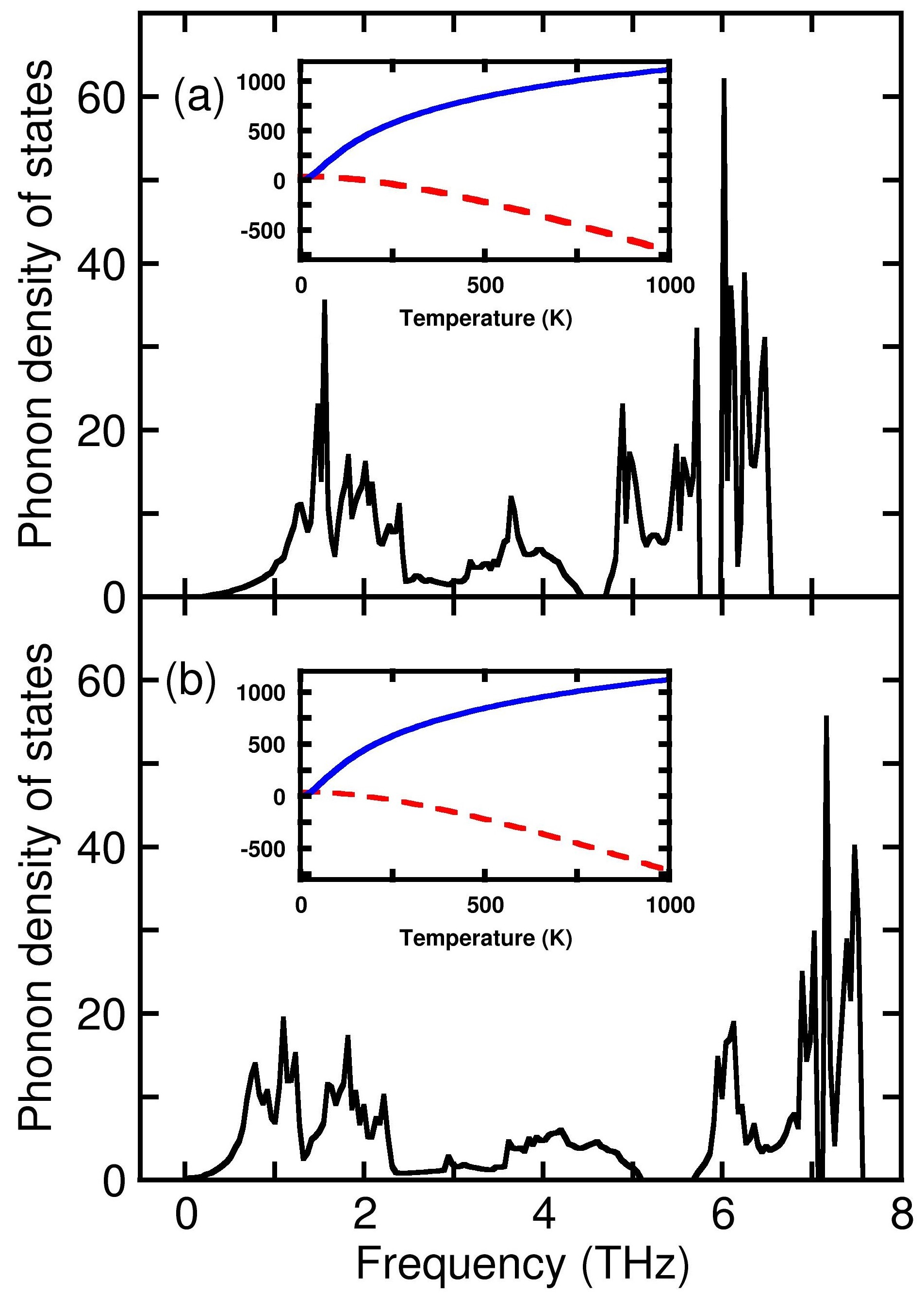}
\caption{(Color online) The calculated phonon density of states for the (a) parent structure and (b) high pressure structure are shown. The inset of the two panels shows the calculated entropy (solid line) in the unit of J/K.mole and the Helmholtz free energy (dotted line) in unit of kJ/mole.}
\label{phonon}
\end{figure}

In order to establish the calculated results the stability of the compound is very important to examine. It becomes very necessary when the material is not yet synthesized experimentally. Therefore, we examine the dynamical stability through the calculated phonon density of states whereas the thermodynamic stability via calculating the Helmholtz free energy and the entropy as shown in Fig. \ref{phonon} . We have calculated the phonon frequencies via finite displacement method using the first principles derived Hessian matrix. The thermodynamic quantities can be obtained through the phonon energy by integrating the phonon frequencies over the entire brillouin zone under harmonic approximation\cite{togo}. 

The calculations show the absence of any imaginary and negative frequency mode in the phonon density of states confirms the dynamical stability of the both parent and high pressure phase of the ZnGeSb$_2$. Moreover, the calculated Helmholtz free energy remains negative in the entire temperature range, whereas, the entropy of the system increases with the temperature, suggesting thermodynamic stability for the both parent ambient phase as well as the high pressure phase.

\section{Projected Band structures for ambient and High pressure phase}
\begin{figure}[t!]
\includegraphics[width=9cm]{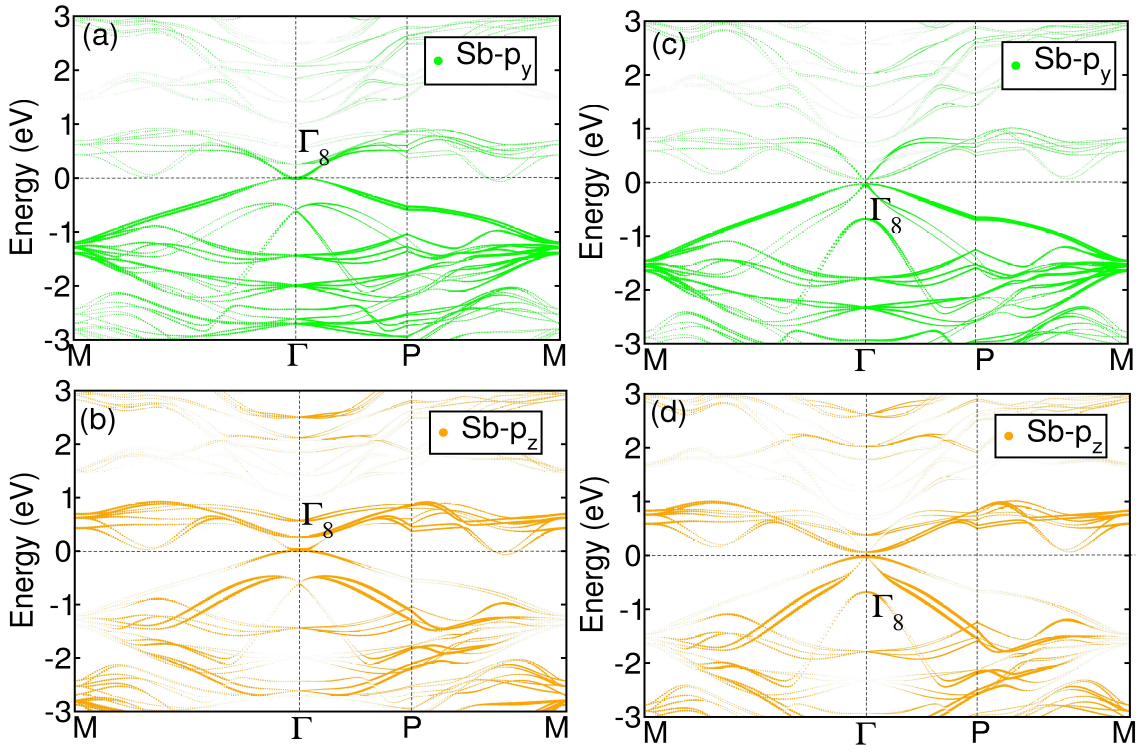}
 \caption{(Color online) (a) and (b) represent ambient (topological non-trivial),  (c) and (d) represent high pressure (topological trivial) phases of ZnGeSb$_2$ respectively for the the projected bands of Sb-$5p_y$, Sb-$5p_z$ respectively. $\Gamma_8$ band characters are marked in the figure and the band inversion in topological phase happens between $5s$ ($\Gamma_6$) and $5p$ ($\Gamma_8$) orbitals at the $\Gamma$ point in the BZ. Fermi energy set at the zero in energy axis.}
\label{fig:app2}
\end{figure}

\begin{figure}[t!]
\includegraphics[width=6cm]{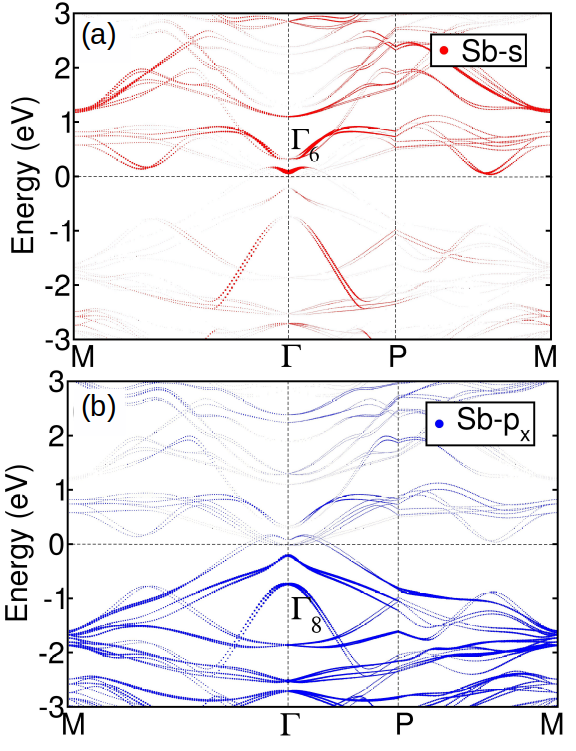}
\caption{(Color online) The orbital projected bands of (a) Sb-$s$ and (b) Sb-$p_x$, for the ZnGeSb$_2$, with the 10$\%$ reduced  $\textit{c}$ lattice parameter, keeping $\textit{a}$ same as that of the equilibrium value of the lattice parameter. Fermi energy set at the zero in energy axis.}
\label{fig:app3} 
\end{figure}

This section shows the projected bands of  Sb-$5p_y$, Sb-$5p_z$ in  ZnGeSb$_2$ at both topological and trivial phase as shown in Fig. \ref{fig:app2}. Figure \ref{fig:app3} ensures that the lattice parameter $\textit{c}$ along the easy axis is a important structural parameter in determining the ground state whether it is in topological or trivial phase.  10$\%$ reduction of lattice parameter $\textit{c}$,  keeping the other structural parameters same as that of the equilibrium value,  ZnGeSb$_2$ switches to trivial phase from its topological ground state.

\section{Wannier charge center calculations and surface states}
\begin{figure}
\center
\includegraphics[width=7cm]{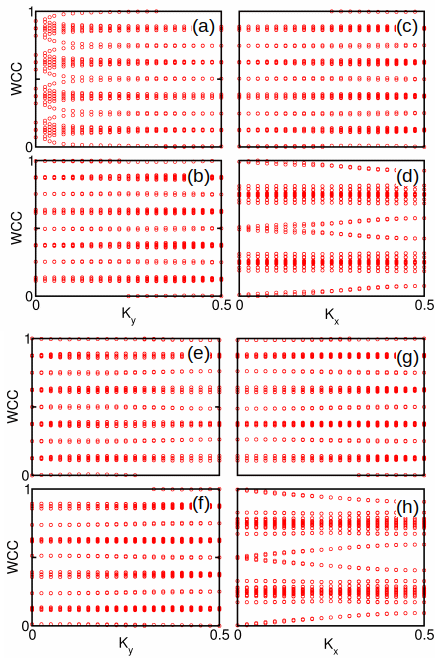}
\caption{(Color online) The flow chart of the average position of Wannier charge centers (WCC) obtained by the Wilson-loop method for the highest occupied valence band for parent state (top panel) and high pressure pressure phase (bottom panel) of ZnGeSb$_2$. (a), (b), (c) and (d) of the each panels represents the calculated WCC for the four time reversal invariant planes $K_x=0.0$ $K_x=0.5$, $K_y=0.5$ and $K_z=0.5$ respectively.}
\label{fig:app4}
\end{figure}

 Here we calculate the topological $Z_2$ invariant for non-centrosymmetric crystal ZnGeSb$_2$ 
using Wannier charge center (WCC) algorithm. The tight binding model has been extracted using maximally 
projected Wannier functions (WFs) for the Zn-3$d$, Ge-3$d$, -4$s$, -4$p$ and Sb-5$p$ orbitals. Here 
we are interested on the highest occupied valence band. The topology comes at $\Gamma$ point where 
band inversion happens from this band for which we are interested to calculate $Z_2$ in ZnGeSb$_2$.
The  WCC are calculated in momentum space by considering four time reversal
invariant planes $K_x=0.5$, $K_y=0.5$, $K_z=0.5$ and $K_z=0.0$ in the three dimensional BZ. For example, the $K_z=0.0$ plane goes through the origin and spanned by reciprocal lattice vector $\bf{G_1}$-$\bf{G_2}$, whereas $K_z=0.5$ plane goes through the $\bf{G_3}$ and spanned by reciprocal lattice vector $\bf{G_1}$-$\bf{G_2}$. 

\begin{figure}
\includegraphics[width=7cm]{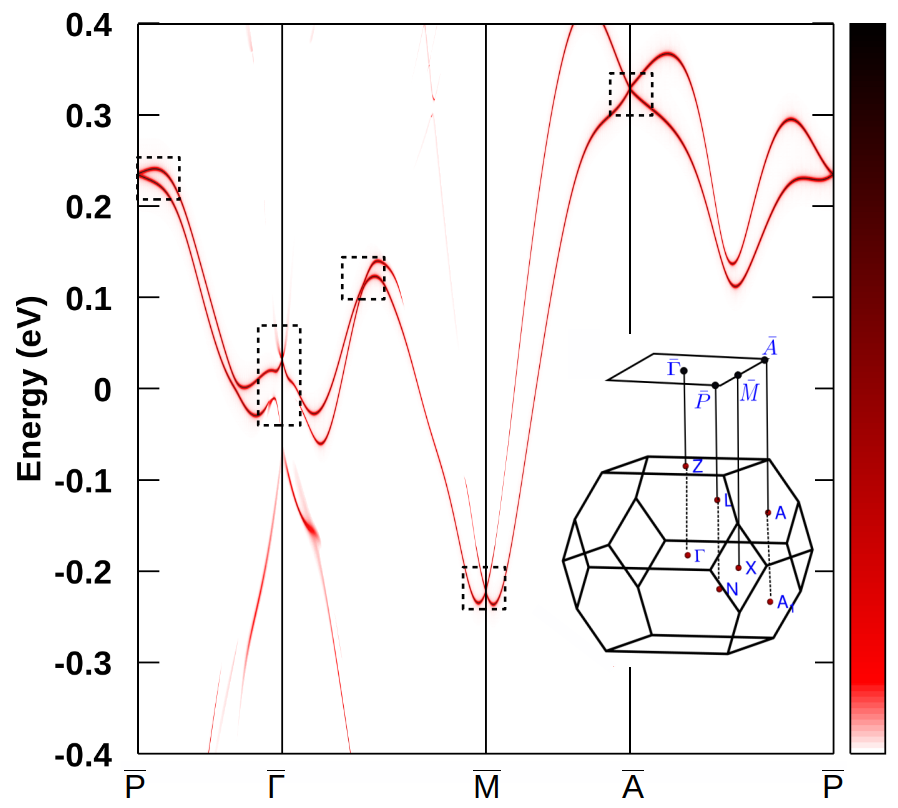}
\caption{Spectral distribution along the surface Time reversal invariant points (TRIM) points for topological (a) non-trivial (ambient pressure) (c) trivial phases (high pressure). (b) TRIM in the BZ in ZnGeSb$_2$ and its surface projection. (d) The enlarged portion of the  spectral distribution near the Fermi energy (set at zero) for the topological phase shows odd number of band crossing.}
\label{fig:app5}
\end{figure}

The calculated spectral distribution at the surface time reversal invariant momenta (TRIM) points in ZnGeSb$_2$ via Green’s function techniques as implemented within PYFPLO. The surface states (SS)
emerging from the conduction band (CB) and valence band (VB) at $\Gamma$ point in the BZ cross an odd number of times (5 times of Dirac crossing) shown in \ref{fig:app5}(a), (d) which ensures the strong topological character. Point to be noted that the crossing are not always appeared at the TRIM points due to the breaking
of inversion symmetry in ZnGeSb 2 structure. We also calculated the spectral distribution at the high pressure
(7.22 GPa), as shown in \ref{fig:app5}(b) where the both surface states are emerging from the conduction band which ensure topologically trivial band insulating phase.

\end{document}